# Aladdin: Augmenting Urban Environments with Local Area Linked Data-Casting

Tom Heath

Talis Systems Ltd,
43 Temple Row, Birmingham, B2 5LS, UK
`tom.heath@talis.com`
`http://www.talis.com/`

**Abstract.** Urban environments are brimming with information sources, yet these are typically disconnected from related information on the Web. Addressing this disconnect requires an infrastructure able to disseminate information to a specific *micro-location*, to be consumed by interested parties. This paper proposes *Aladdin*, an infrastructure for highly localised broadcast of Linked Data via radio waves. When combined with data retrieved from the Web, Aladdin can enable a new generation of micro-location-aware mobile applications and services.

## 1 Introduction

Urban environments are brimming with information sources – street signs, public notices, boards advertising offers in shops and restaurants – which are a key feature of our existence within these spaces. In parallel, the Web is home to immensely rich data sets that describe these urban environments, and services that provide notification of time-sensitive offers and services in a particular locality. Currently, however, the modes of consumption of these information sources is largely disjoint, with the former consumed as we navigate through physical space, and the latter consumed through various windows onto cyberspace.

Applications such as *Foursquare*[1] attempt to bridge these worlds by exploiting the location-awareness of smartphones, yet already suffer from information overload due to the volume of businesses and points of interest situated in densely populated areas. This is compounded by imprecise data about a smartphone user's whereabouts. Location-finding based on cell-tower triangulation can function indoors but has low precision, while GPS is far more precise but is compromised within buildings or in built-up areas. The result is a deluge of information, only some of which is relevant to the user's precise location.

As competition increases for the attention of individuals in a particular location, there is growing need for an infrastructure for delivering data to users in a specific *micro-location* (i.e. a resolution of a few metres), indoors and out. This paper proposes *Aladdin*, an infrastructure for short range radio broadcast of small data packets. By using this principle of *datacasting* to deliver Linked Data [2] payloads, a new generation of interactive and highly-localised services can be delivered to users, thereby augmenting our experience of urban environments.

---
[1] `https://foursquare.com/`

## 2   Aladdin: A User Scenario

Zach is accompanying Amy to Birmingham while she attends an industry conference. He decides to spend the day exploring the city. As he leaves the hotel his Aladdin-enabled smartphone picks up a broadcast message sent from a lamppost across the street – it's from the local tourist information service and is designed to help visitors find their way around. Zach's smartphone interprets the message, retrieves some additional data from the Web, and gives him a choice of various zones of the city to head for. Today doesn't feel like a shopping day, so Zach selects "Birmingham Museum" from the "Museums and Galleries" section and his smartphone points him in the right direction. As Zach tours the museum his smartphone picks up broadcast messages from each exhibit, giving links to an audio commentary from the Web and additional historical background.

Suffering from museum fatigue, Zach heads outside in search of some lunch, where his smartphone receives a number of broadcast messages from restaurants across the square, advertising their cuisine, prices and special offers. *Luigi's* has a three course Italian lunch deal at a very reasonable price, so Zach's smartphone cross-references this with reviews from the Web and confirms it's a good choice. Once inside, Zach's smartphone picks up the broadcast message containing a link to the Luigi's menu, which the phone overlays with feedback from the Web about particular dishes. Later, as he strolls back to the hotel, Zach wonders how to spend the evening – there's a limit to how long he can stand making small talk with Amy's colleagues. By chance he passes the independent cinema, which is broadcasting a list of what's showing tonight – all obscure titles that Zach's never heard of. Fortunately, the smartphone is able to retrieve summaries of each, which he can peruse at his leisure before heading out for the evening.

## 3   The Aladdin Architecture and Technical Approach

### 3.1   Overview

The key components in the Aladdin architecture are *beacons* and *readers*. Beacons are simple low-power radio transmitters that broadcast data packets containing Linked Data payloads, encoded in RDF. A reader within range of the broadcast receives the data packet sent by the beacon, interprets the payload, and may trigger other actions or applications depending on rules defined by the device owner.

### 3.2 Specifics

Beacons would typically be installed on external or internal walls of buildings, street furniture (e.g. lampposts, signposts, bus stops) or vehicles (e.g. buses, trains, taxis). Those beacons whose broadcast messages change rarely could be implemented on cheaper write-once hardware (with solar power where conditions permit), while those whose content changes regularly could support remote updates over a conventional network connection. For regulatory reasons, and to exploit economies of scale from existing hardware production, the unlicensed 2.4-2.5GHz frequency range may be best suited to Aladdin broadcast messages.

The most common form of reader in the Aladdin scenario would be a smartphone, however the concept is equally applicable to other moving objects where data services may be of value (e.g. cars). Readers may need to consume many broadcast messages in quick succession. Therefore, payloads should be small and contain just enough data to convey the nature of the message, plus pointers to additional data that may be retrieved over conventional data networks.

By definition datacast is a broadcast medium. This unidirectional mode of communication helps guarantee a basic level of privacy for Aladdin users, which in turn broadens the range of potential applications of the medium.

## 4 Applications of Aladdin

The applications of Aladdin are numerous. Beacons could augment or replace existing street signage or be used to notify residents of forthcoming roadworks or other disruptions. Aladdin-enabled bus stops would no longer need to be equipped with expensive display screens showing timetables or delays as pointers to this information could all be delivered via datacasting. As in the scenario above, museums and galleries could use beacons to enhance the visitor experience, while Aladdin provides an infrastructure through which restaurants, bars and shops could reach out to potential customers.

Once indoors, the possibilities increase further: restaurants could broadcast messages with links to the current menu, including the nutritional value, allergenic contents, and origin of the key ingredients, while local authorities could augment each street with the results of hygiene inspections of each establishment!

On a personal level, beacons in taxis could broadcast the certification details of the driver, which a smartphone app could automatically verify via a Web service, alerting the user if the details don't check out or logging the details remotely as a personal safety measure. Applications could even arise allowing smartphone users to run temporary beacons for dating or social networking.

## 5 Related Work

Augmenting environments through mobile devices and Web technologies is not new. For example, [1] and [3] describe prototypes for personalised museum experiences that exploit mobile and Web technologies. More generally, many location-based apps enable users to form connections with specific establishments or locations, e.g. through *checking-in*. However, these services naturally constrain who can participate and the information that can be disseminated to users; i.e. the platform is not open. In summary, these prototypes and services suffer from the same limitation, namely the lack of a universal, open and generic infrastructure for engaging with people in a very specific location.

From a technical perspective, QR codes[2] are increasingly used to associate physical locations with Web addresses from which related information can be retrieved. The shortcoming of this approach is that a mobile device carried in a pocket can not simply consume QR codes as the owner passes them. RFID tags may overcome this issue, but are typically coupled to proprietary identification schemes (e.g. EPC[3]) that are administered by a central authority, and require access to proprietary databases to be looked up. This introduces potential barriers to widespread uptake. Aladdin, by contrast is based on the use of open standards already well established and widely deployed.

## 6   Conclusions and Research Challenges

This paper has presented the concept of Aladdin, a generic, open infrastructure for radio broadcast of Linked Data to users in a very specific location. This vision presents many opportunities for compelling applications, as well as thorny research issues. Through adoption of the well-established Semantic Web technology stack, the Aladdin proposal overcomes many challenges related to knowledge representation and data exchange. However, further multi-disciplinary effort is required to demonstrate datacasting with an RDF/Linked Data payload, particularly within a specific micro-location. In addition, the notion of an urban environment with many beacons broadcasting messages offers many opportunities for misuse, with *Aladdin spam* a conceivable possibility. These issues do not negate the value of the idea. On the contrary, this use case can drive forward further research into key issues (such as authenticity, validity, verification, and trust) that are of critical importance to the Semantic Web as a whole.

---

[2] http://en.wikipedia.org/wiki/QR_code
[3] http://www.gs1.org/epcglobal